\documentclass[12pt,draftclsnofoot,onecolumn,comsoc]{IEEEtran}
\usepackage{amsmath,amsfonts}
\usepackage{array}
\usepackage{cite}
\usepackage{algorithm}  
\usepackage{algorithmic}
\usepackage{subfigure}
\usepackage[caption=false,font=normalsize,labelfont=sf,textfont=sf]{subfig}
\usepackage{textcomp}
\usepackage{stfloats}
\usepackage{url}
\usepackage{verbatim}
\usepackage{graphicx}
\usepackage{amssymb}
\usepackage{color}
\usepackage{graphics}
\usepackage{epstopdf}
\newtheorem{lemma}{\underline{Lemma}}%[section]

\newtheorem{proposition}{\underline{Proposition}}%[section]
\usepackage{balance}

\begin{document}

\title{Optimized Design for IRS-Assisted Integrated Sensing and Communication Systems in \\ Clutter Environments}

\author{Chikun Liao,~\IEEEmembership{Student Member, IEEE}, Feng Wang,~\IEEEmembership{Member, IEEE}, \\and Vincent K. N. Lau,~\IEEEmembership{Fellow, IEEE}

\thanks{C. Liao and F. Wang are with the School of Information Engineering, Guangdong University of Technology, Guangzhou 510006, China (e-mail: fengwang13@gdut.edu.cn).}
\thanks{V. K. N. Lau is with the Department of Electronic and Computer Engineering, The Hong Kong University of Science and Technology, Hong Kong (e-mail: eeknlau@ust.hk).}

\vspace{-1.8cm}
}

\maketitle
\begin{abstract}
In this paper, we investigate an intelligent reflecting surface (IRS)-assisted integrated sensing and communication (ISAC) system design in a clutter environment. Assisted by an IRS equipped with a uniform linear array (ULA), a multi-antenna base station (BS) is targeted for communicating with multiple communication users (CUs) and sensing multiple targets simultaneously. We consider the IRS-assisted ISAC design in the case with Type-I or Type-II CUs, where each Type-I and Type-II CU can and cannot cancel the interference from sensing signals, respectively. In particular, we aim to maximize the minimum sensing beampattern gain among multiple targets, by jointly optimizing the BS transmit beamforming vectors and the IRS phase shifting matrix, subject to the signal-to-interference-plus-noise ratio (SINR) constraint for each Type-I/Type-II CU, the interference power constraint per clutter, the transmission power constraint at the BS, and the cross-correlation pattern constraint. Due to the coupling of the BS's transmit design variables and the IRS's phase shifting matrix, the formulated max-min IRS-assisted ISAC design problem in the case with Type-I/Type-II CUs is highly non-convex. As such, we propose an efficient algorithm based on the alternating-optimization and semi-definite relaxation (SDR) techniques. In the case with Type-I CUs, we show that the dedicated sensing signal at the BS is always beneficial to improve the sensing performance. By contrast, the dedicated sensing signal at the BS is not required in the case with Type-II CUs. Numerical results are provided to show that the proposed IRS-assisted ISAC design schemes achieve a significant gain over the existing benchmark schemes.
\end{abstract}


\section{Introduction}
 The advancement of the beyond-fifth-generation (B5G) and sixth-generation (6G) communication networks has led to various integrated sensing and communication (ISAC) applications, such as auto-driving, vehicle-to-everything, smart home, virtual/augmented reality, and edge intelligence \cite{r1,r2,r3,r4,r5}. In ISAC systems, the sensing and communication capabilities are tightly integrated to provide both sensing and communication services \cite{r6,r7,r8,r9,r10,r11,r12,r13,r14,r15,r16}. The resources of communication and radar sensing can be shared in ISAC systems, thereby boosting the efficiency of utilization in terms of spectrum, hardware, energy, and cost \cite{r17}.

 The emerging intelligent reflective surface (IRS) (also known as reconfigurable intelligent surface (RIS)) has recently attracted a growing interest from academia and industry as a promising approach to improving the communication performance\cite{r18,r19,r20}. Motivated by the effectiveness of IRS-assisted communication systems, the IRS technology has been proposed to assist ISAC systems to further improve the system performance\cite{r21,r22,r23}. For example, the authors in \cite{r21} employed the IRS to mitigate  multi-user interference to improve the communication performance, while matching the radar waveform. The IRS-assisted communication with multiple sensing targets were studied in the mmWave systems \cite{r22}. In \cite{r23} the authors adopted the  Cram$\acute{e}$r-Rao bound (CRB) as a radar sensing performance indicator, and studied the joint waveform and discrete phase shift designs for the IRS-assisted ISAC systems.
 
 Notice that the radar detection and sensing performance relies highly on the direct sensing link from the base station (BS). It is critical to improve the ISAC performance for the sensing targets in the non-line-of-sight (NLoS) areas \cite{r24,r25,r26,r27,r28,r29}. By reconfiguring the wireless propagation environment, the IRS technology can help improve wireless communication rate, as well as to enhance the sensing accuracy and resolution performance \cite{r24}. Specifically, the IRS can constructively enhance the reflected sensing signals and mitigate the malicious interference. Also, the IRS technology can provide additional degrees of freedom to combat the severe channel fading effect. The authors in \cite{r25} considered an IRS-assisted ISAC system, by leveraging IRS to create new channel links to achieve a high-quality communication and sensing performance. The work in \cite{r26} studied a hybrid IRS model comprising active and passive elements to assist the radar and communication, and aimed to maximize the worst-case target illumination power. The authors in \cite{r27} studied using multiple IRSs for further enhancing the ISAC performance. Note that by minimizing the cross-correlation coefficient between radar signals, one can achieve a better radar detection performance \cite{r28}. The authors in \cite{r29} studied the IRS-assisted ISAC system design under the radar signal cross-correlation pattern constraints, where the IRS may introduce additional interference for sensing. Note that the aforementioned works\cite{r24,r25,r26,r27,r28,r29} all considered clutter-free environments, and there lacks investigation of IRS-assisted ISAC designs in a clutter environment.
 
 In a clutter environment, the radar sensing performance is often affected by various clutters, such as trees, tall buildings, and cars\cite{r30}. The reflected signal power from the clutter may be higher than the target's reflected signal power, which can severely degrade the target detection reliability. How to detect and sense multiple targets in a clutter environment is a challenging issue for ISAC system designs\cite{r31,r32,r33}. The clutter can be treated as a signal-dependent type of interference, which depends on a deterministic transmitted signal \cite{r34}. As such, one can adopt the clutter's prior knowledge such as the clutter covariance matrix to achieve the suppression of clutter\cite{r35}. By considering multiple clutters in an IRS-assisted ISAC system, the authors in \cite{r36} aimed to maximize the radar SINR subject to the communication quality of service requirement constraint. Note that the impact of clutter on the blocking of the line-of-sight (LoS) link is ignored in \cite{r36}. The work in \cite{r30} studied the radar design with multiple IRSs in order to assist the sensing services in a clutter environment, where the effective sensing links were created by the IRSs for addressing the unavailability of the direct sensing link. In a clutter environment, the clutter interference to the radar system must be mitigated to achieve a desirable target sensing performance, which motivates our work in this paper.
 
 In this paper, we consider an IRS-assisted ISAC system in a clutter environment. The system consists of one multi-antenna BS providing ISAC service, an IRS with a uniform linear array (ULA), multiple CUs, and multiple targets to be sensed. We consider two types of CUs, i.e., Type-I and Type-II CUs, where the Type-I CU has the ability to eliminate the interference caused by sensing signals, and the Type-II CU cannot eliminate the interference power caused by the sensing signals. In a clutter environment, we pursue an efficient IRS-assisted ISAC design to maximize the minimum sensing beampattern gain among multiple targets. The main contributions of this paper are summarized as follows.

\begin{itemize}
 \item First, we propose an optimization framework in a clutter environment for the IRS-assisted ISAC system design with Type-I or Type-II CUs. Specifically, we maximize the minimum sensing beampattern gain among multiple targets, subject to the BS's transmission power constraint, the signal-to-interference-plus-noise-ratio (SINR) for each Type-I or Type-II CU, the power constraint of each clutter, and the radar signals cross-correlation pattern constraint. We jointly optimize the BS's transmit beamforming vectors for multiple CUs and sensing covariance matrix for multiple targets, as well as the IRS's phase shifting matrix. Due to the variable coupling, the formulated max-min IRS-assisted ISAC design problem in the case with Type-I or Type-II CUs is non-convex, which is difficult to obtain the global solution.
 
 \item Second, by combining the alternating-optimization and semi-definite relaxation (SDR) techniques, we propose a low-complexity method to obtain the solutions for the IRS-assisted ISAC designs in the cases with Type-I and Type-II CUs, respectively. The tightness of the SDR and the computational complexities for the obtained solutions are analyzed. It is shown that the dedicated sensing signal is always required at the BS to enhance the sensing performance in the case with Type-I CUs in clutter environments. By contrast, in the case with Type-II CUs, the dedicated sensing signals can be removed at the BS without loss of the system sensing performance.
  
 \item Finally, extensive numerical results are provided to show the effectiveness of our proposed max-min IRS-assisted ISAC design schemes. In the case with Type-I CUs or Type-II CUs, the proposed design scheme is shown to achieve a significant performance gain over the existing benchmark schemes. It is also shown that the IRS-assisted ISAC system with Type-I CUs always outperform that with Type-II CUs. Furthermore, the deployment of the IRS can effectively help increase the robustness against the link blockage in clutter environments.
\end{itemize}

The rest of this paper is organized as follows. Section II presents the IRS-assisted ISAC system model in a clutter environment and formulates the max-min design problems under consideration. Section III and Section IV propose low-complexity design solutions in the cases with Type-I CUs and Type-II CUs, respectively. Section V shows numerical results to evaluate the proposed design schemes. Finally, we conclude the paper in Section VI.

{\em Notations:} Boldface letters refer to vectors (lower case) or matrices (upper case); rank($\cdot$), tr($\cdot$), diag($\cdot$), and vec($\cdot$) denote the rank, the trace, the diagonal, and the vectorization operations, respectively; $[\cdot]^{\ast}$, $[\cdot]^T$, and $[\cdot]^H$ denote the conjugate, transpose, and Hermitian transpose operations, respectively; $\left\|\cdot\right\|$ denotes the Euclidean norm.

\section{System Model and Problem Formulation}

\begin{figure}
    \centering
    \includegraphics[width=12cm]{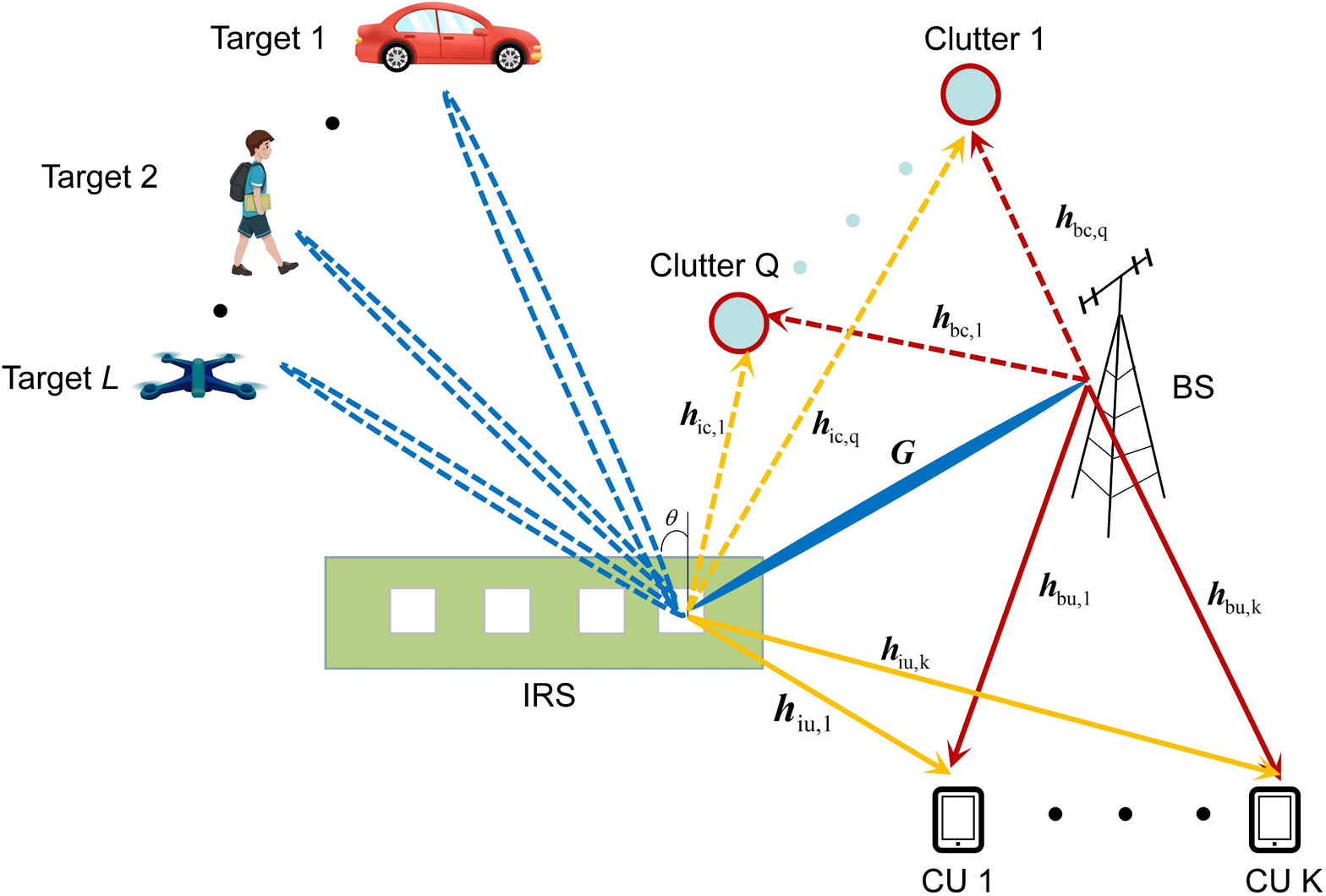}
    \caption{The IRS-assisted ISAC system model in a clutter environment, where no LoS links exist between the BS and multiple targets due to multiple clutters.}\label{fig1}
\end{figure}

 As shown in Fig.~\ref{fig1}, we consider an IRS-assisted ISAC system in a clutter environment, where a multi-antenna BS is assisted by an IRS for sending unicast communication signals for a number of $K$ single-antenna CUs and dedicated radar sensing signals for a total of $L$ targets, and the IRS is deployed to help propagate communication and radar sensing signals. The BS is equipped with $M>1$ antennas, and the IRS is equipped with an uniform linear array (ULA) of $N>1$ passive reflecting elements. We consider a number of $Q$ clutters between the BS and the $L$ sensing targets. The CU set and the clutter set are denoted by ${\cal K}\triangleq \{1,...,K\}$ and $\mathcal{Q}\triangleq \{1,...,Q\}$, respectively. Denote by ${\cal L}\triangleq\{1,...,L\}$ the set of $L$ sensing targets. Due to the signal blockage in propagation, we assume that there exists no line-of-sight (LoS) link between the BS and the $L$ targets for sensing due to the existence  of the $Q$ clutters. We consider that the BS has obtained the global channel state information (CSI) and the location information of the sensing targets during the detection phase. 
 
 Denote by $s_k\in\mathbb{C}$ and $\boldsymbol{w}_k\in\mathbb{C}^{M\times 1}$ the information symbol and transmit beamforming vector for CU $k\in{\cal K}$, respectively. Without loss of generality, we assume that the transmit symbols $\{s_k\}$ are independent and identical distributed (i.i.d.) random variables with zero mean and unit variance, i.e., $s_k \sim \mathcal{CN}(0,1)$, $k\in{\cal K}$. Let $\boldsymbol{x}_0\in \mathbb{C}^{M\times1}$ be the dedicated sensing signal of the BS with zero mean, i.e., $\mathbb{E}[\boldsymbol{x}_0]=0$, and the sensing covariance matrix $\boldsymbol{R}_0=\mathbb{E}\left[\boldsymbol{x}_0\boldsymbol{x}_0^H\right]$ is with a general rank, i.e., $\text{rank}(\boldsymbol{R}_0)\geq 1$. As a result, the ISAC transmit signal $\boldsymbol{x}\in\mathbb{C}^{M\times1}$ of the BS is modeled as
 \begin{equation}\label{eq.x}
    \boldsymbol{x} = \underbrace{\boldsymbol{x}_0}_{\text{sensing~signal~for}~L~\text{targets}} + \sum_{k=1}^{K}\underbrace{\boldsymbol{w}_k {s}_k}_{\text{commun.~signal for CU}~k}.
 \end{equation}
 
 Based on \eqref{eq.x}, the transmit power constraint of the BS is expressed as 
 \begin{equation}
    \mathbb{E}[\|\boldsymbol{x}\|^2] =  \text{tr}(\boldsymbol{R}_0) + \sum_{k=1}^{K}\left\|\boldsymbol{w}_k\right\|^2 \leq {P_0},
 \end{equation}
 where the expectation $\mathbb{E}[\cdot]$ is taken on the randomness of $\{\boldsymbol{x}_0,s_k\}$, and $P_0$ denotes the maximum transmit power budget of the BS.

\subsection{IRS-assisted Communication Model for Type-I and Type-II CUs}
 Let $\boldsymbol{G}\in \mathbb{C}^{N\times{M}}$ denote the complex-valued channel matrix from the BS to the IRS, and let $\boldsymbol{h}_{bu,k}\in \mathbb{C}^{M\times{1}}$ and $\boldsymbol{h}_{iu,k}\in \mathbb{C}^{N\times{1}}$ denote the complex-valued channel vectors from the BS and the IRS to CU $k$, respectively. Denote by $\phi_n\in{\left(0,2\pi\right]}$ the phase shift value of each passive reflecting element $n\in {\cal N}\triangleq \{1,...,N\}$ at the IRS. Let $\boldsymbol{\Phi}=\text{diag}(e^{j\phi_1},\cdot\cdot\cdot,e^{j\phi_N})\in\mathbb{C}^{N\times N}$ denote the IRS's phase shifting matrix, where $j=\sqrt{-1}$ denotes the imaginary unit. In the IRS-assisted ISAC system under consideration, each CU $k\in{\cal K}$ receives the BS's signal via both the direct BS-CU link and the cascaded BS-IRS-CU link. Accordingly, the received signal $y_k\in\mathbb{C}$ of CU $k\in{\cal K}$ is modeled as
 \begin{subequations}
 \begin{align}
   y_k &= \boldsymbol{h}^H_{\text{CU},k} \boldsymbol{x} + n_k\\
   &= \boldsymbol{h}_{\text{CU},k}^H \Big(\boldsymbol{x}_0 + \sum_{k=1}^{K}\boldsymbol{w}_k s_k \Big) + n_k,~\forall k\in{\cal K},
   \end{align}
 \end{subequations}
 where $\boldsymbol{h}_{\text{CU},k} \triangleq \boldsymbol{G}^H\boldsymbol{\Phi}^H \boldsymbol{h}_{iu,k} + \boldsymbol{h}_{bu,k}$ is defined as the {\em effective} channel vector from the BS to CU $k\in \mathcal{K}$, and $n_k\sim \mathcal{CN}(0,\sigma_k^2)$ denotes the additive white Gaussian noise (AWGN) at CU $k$'s receiver. 

 Depending on the ability of cancelling the sensing signal $\boldsymbol{x}_0$ or not, we consider two types of CUs, i.e., Type-I and Type-II CUs. In the case with Type-I CUs, the sensing signal $\boldsymbol{x}_0$ is globally known, and each Type-I CU can completely cancel the interference generated by the sensing signal $\boldsymbol{x}_0$. Denote by $\gamma^{\text{I}}$ the SINR of Type-I CU $k\in{\cal K}$. Accordingly, we have
 \begin{equation}
    \gamma_{k}^{\text{I}} =  \frac{\left|\boldsymbol{h}^H_{\text{CU},k} \boldsymbol{w}_k\right|^2}{\sum_{j=1,j\neq k}^K\left|\boldsymbol{h}^H_{\text{CU},k} \boldsymbol{w}_j\right|^2 + \sigma_k^2}, ~\forall k\in{\cal K}.
\end{equation}

In the case with Type-II CUs, the sensing signal $\boldsymbol{x}_0$ is {\em not} globally known, and each Type-II CU cannot eliminate the interference power caused by the sensing signal $\boldsymbol{x}_0$. Denote by $\gamma^{\text{II}}$ the SINR of Type-II CU $k\in{\cal K}$. Then, we have
\begin{equation}
    \gamma_{k}^{\text{II}} =  \frac{\left|\boldsymbol{h}_{\text{CU},k}^H \boldsymbol{w}_k\right|^2}{\sum_{j=1,j\neq k}^K\left|\boldsymbol{h}_{\text{CU},k} ^H\boldsymbol{w}_j\right|^2+\left|\boldsymbol{h}_{\text{CU},k}^H \boldsymbol{x}_0\right|^2 + \sigma_k^2},~ \forall k\in{\cal K}.
\end{equation}

\subsection{IRS-assisted Sensing Model for $L$ Targets}
 In the IRS-assisted ISAC system, the IRS can reconfigure a sensing link (i.e., the BS-IRS-target cascaded channel link) to provide a direct target sensing service by properly adjusting the phases of its reflecting elements. Let $\theta_l\in(-\frac{\pi}{2},\frac{\pi}{2})$ denote the angle of target $l\in{\cal L}$ with respect to the IRS. Let $d_{\text{IRS}}$ denote the distance between two consecutive reflecting elements of the IRS, and let $\lambda$ denote the wavelength of the BS's radio-frequency (RF) transmit signal. For each target $l\in{\cal L}$, the array steering vector of the IRS with angle of departure (AoD) $\theta_l$ is given by
\begin{equation}
    \boldsymbol{\alpha}(\theta_l)=\left[1,e^{j\frac{2\pi d_{\text{IRS}}}{\lambda}sin\theta_l},...,e^{j\frac{2\pi(N-1) d_{\text{IRS}}}{\lambda}sin\theta_l}\right]^T,~\forall l\in{\cal L}.
\end{equation}

 In ISAC systems, both the communication signals $\{s_k\}_{k\in{\cal K}}$ and the dedicated sensing signal $\boldsymbol{x}_0$ can be used to illuminate the $L$ targets to satisfy the required sensing beampattern gain constraint. As a result, the beampattern gain $\rho(\theta_l)$ for target $l\in{\cal L}$ located at the AoD angle $\theta_l$ with respect to the IRS is given as \cite{r37}
 \begin{subequations}
 \begin{align}
        \rho(\theta_l)&=\mathbb{E}\Bigg[\Big|\boldsymbol{h}_{\text{target},l}^H\Big(\boldsymbol{x}_0 + \sum_{k=1}^{K}\boldsymbol{w}_k s_k \Big)\Big|^2\Bigg]\\
       &=\boldsymbol{h}_{\text{target},l}^H\Big(\boldsymbol{R}_0 + \sum_{k=1}^K\boldsymbol{w}_k\boldsymbol{w}_k^H\Big)\boldsymbol{h}_{\text{target},l},~\forall l\in{\cal L},
 \end{align}
 \end{subequations}
 where $\boldsymbol{h}_{\text{target},l} \triangleq \boldsymbol{G}^H\boldsymbol{\Phi}^H\boldsymbol{\alpha}(\theta_l)$ is defined as the {\em effective} LoS channel vector from the BS to target $l\in{\cal L}$.

 Furthermore, we consider the cross-correlation pattern constraint for the ISAC signal to sense the $L$ targets with different angles. Let $P^{\text{cross}}$ denote the mean-squared cross-correlation pattern for the $L$ targets. As in \cite{r38}, we have
 \begin{equation}
    P^{\text{cross}} = \frac{2}{L(L-1)}\sum_{l=1}^{L-1}\sum_{i=l+1}^{L}P(\theta_l, \theta_i),
 \end{equation}
 where $P(\theta_l, \theta_i)$ denotes the cross-correlation coefficient between target $l$ at angle $\theta_l$ and another target $i$ at angle $\theta_i$, and we have
 \begin{align}
 P(\theta_l, \theta_i) \triangleq \boldsymbol{h}_{\text{target},l}^H\Big(\boldsymbol{R}_0 + \sum_{k=1}^K\boldsymbol{w}_k\boldsymbol{w}_k^H\Big)\boldsymbol{h}_{\text{target},i},
 \end{align}
 where $l\neq i \in{\cal L}$.

\subsection{Clutter Model for $Q$ Clutters}
 For each clutter $q\in{\cal Q}$, we consider the associated interference power. Let $\boldsymbol{g}_{bc,q}\in\mathbb{C}^{M\times1}$ denote the complex-valued channel vector from the BS to clutter $q\in{\cal Q}$, and let $\boldsymbol{g}_{ic,q}\in \mathbb{C}^{N\times1}$ denote the complex-valued cascaded channel vector from the BS to clutter $q$ via the IRS. As a result, the interference power incurred by each clutter $q\in{\cal Q}$ is expressed as
\begin{subequations}
    \begin{align}
    P^{\text{clutter}}_q &= \mathbb{E}\Bigg[\Big|\boldsymbol{g}_{\text{clutter},q}^H \Big(\boldsymbol{x}_0+\sum_{k=1}^{K}\boldsymbol{w}_k s_k\Big)\Big|^2\Bigg]\\
    &=\boldsymbol{g}_{\text{clutter},q}^H \Big(\boldsymbol{R}_0+\sum_{k=1}^K\boldsymbol{w}_k\boldsymbol{w}_k^H\Big)\boldsymbol{g}_{\text{clutter},q},~\forall q\in{\cal Q},
    \end{align}
\end{subequations}
where $\boldsymbol{g}_{\text{clutter},q} \triangleq \boldsymbol{G}^H\boldsymbol{\Phi}^H\boldsymbol{g}_{ic,q} + \boldsymbol{g}_{bc,q}$ is defined as the {\em effective} channel vector from the BS to clutter $q\in \cal{Q}$.
%where $\eta_q$ denotes the maximum tolerable power for clutter $q$.

\subsection{Problem Formulation}
 In this paper, we are interested in maximizing the minimum sensing beampattern gain among the $L$ targets, by jointly optimizing the BS's transmit beamforming vectors $\{\boldsymbol{w}_k\}_{k\in{\cal K}}$ for communication and covariance matrix $\boldsymbol{R}_0$ for sensing, as well as the IRS's phase shifting matrix $\boldsymbol{\Phi}$. The constraints include the BS's transmit power constraint, the SINR constraint for each Type-I or Type-II CU $k\in{\cal K}$, the interference power constraint for each clutter $q\in{\cal Q}$, and the mean-squared cross-correlation pattern constraint for the $L$ targets. 
 
 In the case with Type-I CU, we formulate the following optimization problem as
 \begin{subequations}\label{eq.p1}
\begin{align}
 (\text{P}1): &\max_{\{\boldsymbol{w}_k\}_{k\in{\cal K}},\boldsymbol{R}_0\succeq \boldsymbol{0},\boldsymbol{\Phi}}~  \min_{l\in \cal{L}} ~\rho(\theta_l)\\
        \text{s.t.}~&\sum_{k=1}^{K}\left\|\boldsymbol{w}_k\right\|^2 + \text{tr}(\boldsymbol{R}_0)\leq P_0 \\
        &\gamma_{k}^{\text{I}} \geq \Gamma_k,~\forall k\in{\cal K}\\
        &P_q^{\text{clutter}}\leq \eta_q,~\forall q\in{\cal Q}\\
        & P^{\text{cross}} \leq \xi,
\end{align}
 \end{subequations}
 where $\Gamma_k$ denotes the SINR threshold prescribed for each Type-I CU $k\in{\cal K}$, $\eta_q$ denotes the given interference power threshold for clutter $q\in{\cal Q}$, and $\xi$ denotes the predefined power threshold for the mean-squared cross-correlation pattern $P^{\text{cross}}$.
 
 In the case with Type-II CU, we have the following design optimization problem as
 \begin{subequations}\label{eq.p2}
 \begin{align}
        (\text{P}2): &\max_{\{\boldsymbol{w}_k\}_{k\in{\cal K}},\boldsymbol{R}_0\succeq \boldsymbol{0},\boldsymbol{\Phi}}~ \min_{l\in \mathcal{L}} ~\rho(\theta_l)\\
        \text{s.t.}~&~\gamma_{k}^{\text{II}} \geq \Gamma_k,~\forall k\in{\cal K}\\
         &~ (\ref{eq.p1}b),~(\ref{eq.p1}d),~{\text{and}}~(\ref{eq.p1}e),\notag 
\end{align}
\end{subequations}
where $\Gamma_k$ denotes the predefined SINR threshold for each Type-II CU $k\in{\cal K}$.

 In this paper, we consider problems (P1) and (P2) are always guaranteed to be feasible, by properly managing the admission of CUs and mitigating interference powers. Note that the BS's transmit beamforming vectors $\{\boldsymbol{w}_k\}_{k\in{\cal K}}$, the BS's sensing covariance matrix $\boldsymbol{R}_0$, and the IRS's phase shifting matrix $\boldsymbol{\Phi}$ are coupled, problems (P1) and (P2) are highly non-convex. It is highly complicated to obtain the global solutions for problems (P1) and (P2). Alternatively, based on the alternating-optimization method, we next pursue to obtain the low-complexity sub-optimal solutions for problems (P1) and (P2).
 
\section{Proposed Solution for (P1) with Type-I CUs}
In this section, we propose an efficient alternating-optimization-based solution for problem (P1) in the case with Type-I CUs, and then discuss the computational complexity. 

\subsection{Optimization of BS's Transmission Design $(\{\boldsymbol{w}_k\}_{k\in{\cal K}},\boldsymbol{R}_0)$ for ISAC}

In this subsection, under the fixed phase shifting matrix $\boldsymbol{\Phi}$ of the IRS, we optimize the BS's transmit beamforming vectors $\{\boldsymbol{w}_k\}_{k\in{\cal K}}$ for the $K$ Type-I CUs and covariance matrix $\boldsymbol{R_0}$ for sensing $L$ targets.

To this end, we introduce a beamforming matrix $\boldsymbol{W}_k\triangleq \boldsymbol{w}_k\boldsymbol{w}_k^H$ for each Type-I CU $k\in{\cal K}$. It is clear that $\boldsymbol{W}_k\succeq \boldsymbol{0}$ and $\text{rank}(\boldsymbol{W}_k)\leq 1$, $\forall k\in{\cal K}$. By substituting $\boldsymbol{w}_k\boldsymbol{w}_k^H$ with $\boldsymbol{W}_k$ for each $k\in{\cal K}$ and fixing the IRS's phase shifting matrix $\boldsymbol{\Phi}$, the original max-min ISAC design problem (P1) in the case with Type-I CUs is equivalently transformed as
\begin{subequations}\label{eq.p1-1}
 \begin{align}
  (\text{P1.1}): &\max_{\{\boldsymbol{W}_k\}_{k\in{\cal K}},\boldsymbol{R}_0\succeq 0}~  \min_{l\in \mathcal{L}}~ \boldsymbol{h}_{\text{target},l}^H\Big(\boldsymbol{R}_0+\sum_{k=1}^K\boldsymbol{W}_k\Big)\boldsymbol{h}_{\text{target},l}\\
  \text{s.t.}~ &~\text{tr}(\boldsymbol{R}_0)+\sum_{k=1}^K\text{tr}(\boldsymbol{W}_k)\leq P_0 \\
  &~\frac{1}{\Gamma_k}\text{tr}(\boldsymbol{h}_{\text{CU},k} \boldsymbol{h}_{\text{CU},k}^H \boldsymbol{W}_k) - \sum_{j=1,j\neq k}^K\text{tr}(\boldsymbol{h}_{\text{CU},k}\boldsymbol{h}_{\text{CU},k}^H\boldsymbol{W}_j)-\sigma_k^2\geq 0,~\forall k \in \mathcal{K}\\
  &~\boldsymbol{g}_{\text{clutter},q}^H\Big(\boldsymbol{R}_0 +  \sum_{k=1}^K\boldsymbol{W}_k \Big) \boldsymbol{g}_{\text{clutter},q} \leq \eta_q,~ \forall q \in \cal{Q}\\
  &~\frac{2}{L^2-L}\sum_{l=1}^{L-1}\sum_{i=l+1}^{L} \boldsymbol{h}_{\text{target},l}^H\Big(\sum_{k=1}^K\boldsymbol{W}_k+\boldsymbol{R}_0\Big)\boldsymbol{h}_{\text{target},i}^H \leq \xi\\
  &~\boldsymbol{W}_k\succeq \boldsymbol{0},~\text{rank}(\boldsymbol{W}_k)\leq 1, \forall k \in \mathcal{K},
\end{align}
\end{subequations}
where the design variables are $\{\boldsymbol{W}_k\}_{k\in{\cal K}}$ and $\boldsymbol{R}_0$. Due to the rank-one constraints in (\ref{eq.p1-1}f), problem (P1.1) is still non-convex. By employing the celebrated SDR technique\cite{r39} for problem (P1.1), we are ready to obtain a convex optimization with respect to $(\{\boldsymbol{W}_k\}_{k\in{\cal K}},\boldsymbol{R}_0)$ by removing the rank-one constraints $\text{rank}(\boldsymbol{W}_k)\leq 1$,$\forall k\in{\cal K}$. As a result, the relaxed problem (P1.1) can be efficiently solved by using standard convex solvers such as the CVX toolbox \cite{r40}. 

Let $(\{\boldsymbol{W}_k^{\text{I}*}\}_{k\in{\cal K}}, \boldsymbol{R}_0^{\text{I}*})$ denote the optimal solution to the relaxed problem (P1.1). In particular, if $\text{rank}(\boldsymbol{W}_k^{\text{I}*})\leq 1$, $\forall k\in{\cal K}$, then the relaxation of problem (P1.1) is tight, which implies that $(\{\boldsymbol{W}_k^{\text{I}*}\}_{k\in{\cal K}}, \boldsymbol{R}_0^{\text{I}*})$ is the optimal solution to problem (P1.1). In this case, by implementing eigenvalue decomposition (EVD) for each $\boldsymbol{W}_k^{\text{I}*}$, we have $\boldsymbol{W}_k^{\text{I}*}=\lambda_k\boldsymbol{v}_k\boldsymbol{v}^{H}_k$, $\forall k\in{\cal K}$, where $\lambda_k>0$ denotes the maximal eigenvalue of $\boldsymbol{W}_k^{\text{I}*}$, and $\boldsymbol{v}_k$ denotes the corresponding unit-norm eigenvector associated with the eigenvalue $\lambda_k$. Then, we are ready to obtain the optimal solution $(\{\sqrt{\lambda}_k\boldsymbol{v}_k\}_{k\in{\cal K}},\boldsymbol{R}_0^{\text{I}*})$ to problem (P1) under the given phase shifting matrix $\boldsymbol{\Phi}$ of the IRS. On the other hand, if there exists $\text{rank}(\boldsymbol{W}_k^{\text{I}*}) > 1$ for a certain CU $k\in{\cal K}$, then a reconstruction procedure for $(\{\boldsymbol{W}_k^{\text{I}*}\}_{k\in{\cal K}}, \boldsymbol{R}_0^{\text{I}*})$ is required to obtain the optimal solution to problem (P1.1) (i.e., problem (P1) under the given $\boldsymbol{\Phi}$), which is stated in the following lemma.

\begin{lemma}\label{Prop.v1}
In the case with $\text{rank}(\boldsymbol{W}_k^{\text{I}*}) > 1$ for a certain $k\in{\cal K}$, the optimal solution $(\{\tilde{\boldsymbol{w}}^{\text{I}*}_k\}_{k\in{\cal K}},\tilde{\boldsymbol{R}}^{\text{I}*}_0)$ to problem (P1) under the fixed phase shifting matrix $\boldsymbol{\Phi}$ of the IRS is obtained as
\begin{subequations}\label{eq.lem1}
\begin{align}
    \tilde{\boldsymbol{w}}^{\text{I}*}_k&= \frac{1}{\sqrt{\boldsymbol{h}^H_{\text{CU},k}\boldsymbol{W}_k^{\text{I}*}\boldsymbol{h}_{\text{CU},k}}} \boldsymbol{W}_k^{\text{I}*}\boldsymbol{h}_{\text{CU},k},~\forall k\in{\cal K} \\
    \tilde{\boldsymbol{R}}^{\text{I}*}_0&=\boldsymbol{R}_0^{\text{I}*}+\sum_{k=1}^{K}\boldsymbol{W}_k^{\text{I}*}- \sum_{k=1}^{K}\tilde{\boldsymbol{w}}^{\text{I}*}_k(\tilde{\boldsymbol{w}}^{\text{I}*}_k)^H.
\end{align}
\end{subequations}
\end{lemma}
\begin{IEEEproof}
See Appendix~\ref{proof-Prop.v1}.
\end{IEEEproof}

\subsection{Optimization of IRS's Phase Shifting Matrix $\boldsymbol{\Phi}$}

In this subsection, under the fixed transmit beamforming vectors $\{\boldsymbol{w}_k\}_{k\in{\cal K}}$ for the $K$ CUs and covariance matrix $\boldsymbol{R}_0$ for the $L$ targets, we optimize the IRS's phase shifting matrix $\boldsymbol{\Phi}$ of problem (P1). 

We first define $\boldsymbol{\phi}\triangleq\text{vec}(\boldsymbol{\Phi})=[e^{j\phi_1},\cdots,e^{j\phi_N}]^T$ as the IRS's phase shifting vector. By letting $\boldsymbol{R}\triangleq \boldsymbol{R}_0+\sum_{k=1}^K \boldsymbol{w}_k\boldsymbol{w}_k^H$, we introduce the following definitions as
\begin{subequations}\label{eq.def1}
\begin{align}
 &\boldsymbol{R}^{\text{target}}(\theta_l)\triangleq \text{diag}(\alpha(\theta_l))^H\boldsymbol{G}\boldsymbol{R}\boldsymbol{G}^H\text{diag}(\alpha(\theta_l)),~\forall l\in{\cal L} \\
 &\boldsymbol{R}_{k,j}^{\text{CU}} \triangleq \text{diag}(\boldsymbol{h}_{iu,k})^H\boldsymbol{G}\boldsymbol{w}_j\boldsymbol{w}_j^H\boldsymbol{G}^H\text{diag}(\boldsymbol{h}_{iu,k}),~\forall k,j\in{\cal K}\\
 &\boldsymbol{R}^{\text{clutter}}_q \triangleq \text{diag}(\boldsymbol{g}_{ic,q})^H\boldsymbol{G}\boldsymbol{R}\boldsymbol{G}^H\text{diag}(\boldsymbol{g}_{ic,q}),~\forall q\in{\cal Q}\\
 &\boldsymbol{b}^{\text{CU}}_{k,j} \triangleq \text{diag}(\boldsymbol{h}_{iu,k})^H\boldsymbol{G}\boldsymbol{w}_j\boldsymbol{w}_j^H\boldsymbol{h}_{bu,k},~\forall k,j\in{\cal K}\\
 &\boldsymbol{b}^{\text{clutter}}_q \triangleq \text{diag}(\boldsymbol{g}_{ic,q})^H\boldsymbol{G}\boldsymbol{R}\boldsymbol{g}_{bc,q},~\forall q\in{\cal Q}.
\end{align}
\end{subequations}
Based on \eqref{eq.def1}, we define the augmented matrices as $\bar{\boldsymbol{R}}^{\text{target}}(\theta_l)=[\boldsymbol{R}^{\text{target}}(\theta_l),0;0,0]$, $\forall l\in{\cal L}$, $\bar{\boldsymbol{R}}_{k,j}^{\text{CU}}=[\boldsymbol{R}_{k,j}^{\text{CU}}, \boldsymbol{b}_{k,j}^{\text{CU}};(\boldsymbol{b}_{k,j}^{\text{CU}})^H,0]$, $\forall k,j\in{\cal K}$, $\bar{\boldsymbol{R}}^{\text{clutter}}_q = [\boldsymbol{R}^{\text{clutter}}_q,\boldsymbol{b}_{q}^{\text{clutter}};(\boldsymbol{b}_{q}^{\text{clutter}})^H,0], \forall q\in{\cal Q}$.

Then, with the change of variable $\boldsymbol{\phi}\triangleq\text{vec}(\boldsymbol{\Phi})$ and by fixing $(\{\boldsymbol{w}_k\}_{k\in{\cal K}},\boldsymbol{R}_0)$, problem (P1) in the case with Type-I CUs is recast as
\begin{subequations}\label{eq.prob12}
 \begin{align}
    (\text{P1.2}):&~\max_{\bar{\boldsymbol{\phi}}} \min_{l\in \mathcal{L}}~ \bar{\boldsymbol{\phi}}^H\bar{\boldsymbol{R}}^{\text{target}}(\theta_l)\bar{\boldsymbol{\phi}}\\
    \text{s.t.} &~\bar{\boldsymbol{\phi}}^H\Big(\frac{1}{\Gamma_k}\bar{\boldsymbol{R}}^{\text{CU}}_{k,k}-\sum_{j=1,j\neq k}^K \bar{\boldsymbol{R}}^{\text{CU}}_{k,j}\Big)\bar{\boldsymbol{\phi}} \geq c_k^{\text{I}},~ \forall k \in \mathcal{K}\\
    &~\bar{\boldsymbol{\phi}}^H\bar{\boldsymbol{R}}^{\text{clutter}}_q\bar{\boldsymbol{\phi}} \leq \eta_q -|\boldsymbol{g}^{H}_{bc,q}\boldsymbol{R}\boldsymbol{g}_{bc,q}|,~\forall q\in \cal{Q}\\
    &~ \frac{2}{L^2-L}\sum_{l=1}^{L-1}\sum_{i=l+1}^{L}\bar{\boldsymbol{\phi}}^H\bar{\boldsymbol{R}}^{\text{target}}_{l,i}\bar{\boldsymbol{\phi}} \leq \xi\\
    &~\boldsymbol{\phi}=[e^{j\phi_1},...,e^{j\phi_N}]^T,~\phi_n\in[0,2\pi],~\forall n\in\{1,...,N\},
\end{align}
\end{subequations}
where $c^{\text{I}}_k\triangleq \sum_{j=1,j\neq k}^K |\boldsymbol{h}^H_{bu,k}\boldsymbol{w}_j|^2 -\frac{1}{\Gamma_k}|\boldsymbol{h}_{bu,k}^H\boldsymbol{w}_k|^2 + \sigma_k^2$, $\forall  k\in{\cal K}$, and $\bar{\boldsymbol{\phi}}=[\boldsymbol{\phi}^H,1]^H$ is the design variable vector. Note that problem (P1.2) is non-convex, due to the non-convexity of both the SINR constraints  (\ref{eq.prob12}b) and the unit-modulus constraints of (\ref{eq.prob12}e). To address this issue, we define $\boldsymbol{V}\triangleq \bar{\boldsymbol{\phi}}\bar{\boldsymbol{\phi}}^H$. By removing the rank-one constraint $\text{rank}(\boldsymbol{V})=1$, we relax problem (P1.2) into the following SDP \cite{r41}: 
\begin{subequations}
 \begin{align}
    (\text{P1.3}):&~\max_{\boldsymbol{V}}~  \min_{l\in \mathcal{L}}~ \text{tr}(\bar{\boldsymbol{R}}^{\text{target}}(\theta_l)\boldsymbol{V})\\
        \text{s.t.}~ &
    \text{tr}\Big(\Big(\frac{1}{\Gamma_k}\bar{\boldsymbol{R}}^{\text{CU}}_{k,k}-\sum_{j=1,j\neq k}^K \bar{\boldsymbol{R}}^{\text{CU}}_{k,j}\Big){\boldsymbol{V}}\Big) \geq c^{\text{I}}_k,~ \forall k \in \mathcal{K}\\
    &\text{tr}\Big(\bar{\boldsymbol{R}}^{\text{clutter}}_q\boldsymbol{V} \Big) \leq \eta_q -|\boldsymbol{g}^{H}_{bc,q}\boldsymbol{R}\boldsymbol{g}_{bc,q}|,~\forall q\in \cal{Q}\\
    & \frac{2}{L^2-L}\sum_{l=1}^{L-1}\sum_{i=l+1}^{L}\text{tr}\Big(\bar{\boldsymbol{R}}^{\text{target}}_{l,i}{\boldsymbol{V}}\Big) \leq \xi\\
    &\boldsymbol{V}\succeq \boldsymbol{0},~\boldsymbol{V}_{n,n}=1,~\forall n\in\{1,...,N+1\},
\end{align}
\end{subequations}
which can be efficiently solved via standard convex solvers. Let $\boldsymbol{V}^*$ denote the optimal solution to problem (P1.3). Note that the solution $\boldsymbol{V}^*$ is not always guaranteed to be of rank-one. In the case with $\text{rank}(\boldsymbol{V}^*)>1$, we employ a Gaussian randomization procedure\cite{r42} to generate a feasible rank-one matrix. We generate Gaussian randomization vectors 
$\boldsymbol{r} \sim \mathcal{CN}(0,\boldsymbol{V}^*)$ and we construct the candidate approximate solution to problem (P1.2) as $\boldsymbol{v}=e^{j\arg([\frac{\boldsymbol{r}}{\boldsymbol{r}_{N+1}}]_{(1:N)})}$, the objective value is approximated as the maximum one among all these random realizations. Thus, the randomization realization number is set to be 5000 to ensure that there are enough times to guarantee that the objective value is non-decreasing over iterations. 

\begin{algorithm}
\caption{Proposed alternating-optimization algorithm for solving (P1)}
    \begin{algorithmic}[1]
    \STATE Initialize the phase shifting matrix $\Phi^{(1)}$ and the iteration index $i = 1$;
        \REPEAT
            \STATE {For given $\boldsymbol{\Phi}^{(i)}$, obtain the optimal solution $\boldsymbol{W}_k^{(i)}$ and $\boldsymbol{R}_0^{(i)}$ to problem (P1.1) based on Proposition~1;}
            \STATE For given $\boldsymbol{W}_k^{(i+1)}$ and $\boldsymbol{R}_0^{(i+1)}$, solve problem (P1.3), and then obtain an approximate rank-one solution $\boldsymbol{\Phi}^{(i+1)}$ based on the Gaussian randomization method;
            \STATE  Set $i = i+1$;
       \UNTIL The objective value between two iterations is smaller than a tolerance level $\epsilon$ or the maximum number of iterations is reached.
    \end{algorithmic}
\end{algorithm}

\subsection{Proposed Algorithm and Complexity Analysis}
In Algorithm~1, we summarize the proposed alternating-optimization-based algorithm for solving the max-min IRS-assisted ISAC design problem (P1) in the case with Type-I CUs. Note that the achieved values of problem (P1.1) and (P1.3) are non-decreasing over the  alternating-optimization-based iterations. The alternation process is repeated continuously until the improvement in the objective value between the two iterations is less than the tolerance level $\epsilon$, and the convergence of the proposed Algorithm~1 is guaranteed.

The computational complexity of the proposed Algorithm 1 is calculated as follows. Denote by $J$ and $I$ the iteration number and the SDR randomization trial number, respectively. With a given tolerable error level $\epsilon$, we solve the max-min problem with the interior-point method \cite{r43} is $\mathcal{O}(J(K^{4.5}N^{4.5})\log(1/\epsilon))$. For each iteration of the Gaussian randomization procedure, it has a computational complexity of $\mathcal{O}(JI(NK+1)^{2})$. As a result, the computational complexity of the proposed Algorithm 1 is $\mathcal{O}(J(K^{4.5}N^{4.5})\log(1/\epsilon)) + \mathcal{O}(JI(NK+1)^{2})$.

\section{Proposed Solution For (P2) with Type-II CUs}

In this section, we propose the alternating-optimization-based scheme for problem (P2) in the case with Type-II CUs.

\subsection{Optimization of BS's Transmission Design $(\{\boldsymbol{w}_k\}_{k\in{\cal K}},\boldsymbol{R}_0)$ for ISAC}

In this subsection, we obtain the BS's beamforming vectors $\{\boldsymbol{w}_k\}_{k\in{\cal K}}$ for the $K$ Type-II CUs and covariance matrix $\boldsymbol{R}_0$ for the $L$ targets.

By introducing $\boldsymbol{W}_k=\boldsymbol{w}_k\boldsymbol{w}_k^H\succeq \boldsymbol{0}$, $\forall k\in{\cal K}$, problem (P2) under the fixed IRS's phase shifting matrix $\boldsymbol{\Phi}$ is expressed as
\begin{subequations}\label{eq.p2-1}
 \begin{align}
  (\text{P2.1}): &\max_{\{\boldsymbol{W}_k\}_{k\in{\cal K}},\boldsymbol{R}_0\succeq 0}~  \min_{l\in \mathcal{L}}~ \boldsymbol{h}_{\text{target},l}^H\Big(\boldsymbol{R}_0+\sum_{k=1}^K\boldsymbol{W}_k\Big)\boldsymbol{h}_{\text{target},l}\\
   \text{s.t.}~ &~\frac{1}{\Gamma_k}\text{tr}(\boldsymbol{h}_{\text{CU},k} \boldsymbol{h}_{\text{CU},k}^H \boldsymbol{W}_k) - \sum_{j=1,j\neq k}^K\text{tr}(\boldsymbol{h}_{\text{CU},k}\boldsymbol{h}_{\text{CU},k}^H(\boldsymbol{W}_j+\boldsymbol{R}_0))-\sigma_k^2\geq 0,~\forall k \in \mathcal{K}\\
   & (\ref{eq.p1-1}b), (\ref{eq.p1-1}d), (\ref{eq.p1-1}e), \text{and}~(\ref{eq.p1-1}f), \nonumber
\end{align}
\end{subequations}
where $\{\boldsymbol{W}_k\}_{k\in{\cal K}}$ and $\boldsymbol{R}_0$ are the optimization variables. Due to the rank-one constraints $\text{rank}(\boldsymbol{W}_k)\leq 1$, $\forall k\in{\cal K}$, in (\ref{eq.p1-1}f), problem (P2.1) is non-convex and difficult to solve. As in the case with Type-I CUs, we employ the SDR method to solve problem (P2.1) by removing the rank-one constraints. 

Let $(\{\boldsymbol{W}^{\text{II}*}_k\}_{k\in{\cal K}}, \boldsymbol{R}^{\text{II}*}_0)$ denote the optimal solution to the relaxed problem (P2.1). If it holds that $\text{rank}(\boldsymbol{W}^{\text{II}*}_k)=1$, $\forall k\in{\cal K}$, then $(\{\boldsymbol{W}^{\text{II*}}_k\}_{k\in{\cal K}}, \boldsymbol{R}^{\text{II}*}_0)$ is the optimal solution to problem (P2.1). In this case, by performing the EVD for each $\boldsymbol{W}^{\text{II}*}_k$, we have $\boldsymbol{W}^{\text{II}*}_k=\lambda_k\boldsymbol{v}_k\boldsymbol{v}^H_k$, $\forall k\in{\cal K}$, where $\lambda_k>0$ is the non-zero eigenvalue of $\boldsymbol{W}^{\text{II}*}_k$ and $\boldsymbol{v}$ is the unit-norm eigenvector associated with the eigenvalue $\lambda_k$. We now obtain the optimal solution $(\{\sqrt{\lambda_k}\boldsymbol{v}_k\}_{k\in{\cal K}},\boldsymbol{R}^{\text{II}*}_0)$ to the original problem (P2) under the given $\boldsymbol{\Phi}$. On the other hand, if there exists $\text{rank}(\boldsymbol{W}^{\text{II}*}_k)>1$ for a certain $k\in{\cal K}$, then we employ the rank-one solution construction procedure as in Lemma~\ref{Prop.v1}. In particular, the obtained solution to problem (P2) under the given $\boldsymbol{\Phi}$ is expressed as $(\{\tilde{\boldsymbol{w}}^{\text{II}*}_k\}_{k\in{\cal K}},\tilde{\boldsymbol{R}}_0^{\text{II}*})$, where $\tilde{\boldsymbol{w}}^{\text{II}*}_k=\frac{1}{\boldsymbol{h}^H_{\text{CU},k}\boldsymbol{W}^{\text{II}*}_k\boldsymbol{h}^H_{\text{CU},k}}\boldsymbol{W}_k^{\text{II}*}\boldsymbol{h}_{\text{CU},k}$, $\forall k\in{\cal K}$, and $\tilde{\boldsymbol{R}}_0^{\text{II}*}=\boldsymbol{R}^{\text{II}*}_0+\sum_{k=1}^K\boldsymbol{W}_k^{\text{II}*}-\sum_{k=1}^K\tilde{\boldsymbol{w}}^{\text{II}*}_k(\tilde{\boldsymbol{w}}^{\text{II}*}_k)^H$.

Note that the optimal solution $(\{\boldsymbol{W}^{\text{II}*}_k\}_{k\in{\cal K}}, \boldsymbol{R}^{\text{II}*}_0)$ to problem (P2.1) is not unique. Interestingly, it is shown that there exists an optimal solution to problem (P2.1) with $\boldsymbol{R}^{\text{II*}}_0=\boldsymbol{0}$, as stated in the following proposition.
\begin{proposition}\label{Prop.v2}
 For problem (P2.1), there always exists an optimal solution $(\{\boldsymbol{W}^{\text{II}*}_k\}_{k\in{\cal K}},\boldsymbol{0})$, i.e., $\boldsymbol{R}^{\text{II}*}_0=\boldsymbol{0}$, which implies that the dedicated sensing signal $x_0$ can be removed at the BS without loss of optimality of problem (P2.1) in the case with Type-II CUs.
\end{proposition}
\begin{IEEEproof}
See Appendix~\ref{proof-Prop.v2}.
\end{IEEEproof}

Note that the feasible region of (P2.1) is included by that of (P1.1). As expected, the beampattern gain achieved by (P1.1) is guaranteed to be greater than that by (P2.1). This is because the dedicated radar signals cannot affect the communication performance of the Type-I CUs, but it is not true for the ISAC system with Type-II CUs. The ISAC system with Type-I CUs is shown to achieve better or equal beampattern gains than the counterpart with Type-II CUs.

%First, the BS is equipped with a ULA device, and the channel matrix from BS to IRS is LoS. Thus $\boldsymbol{G} = \boldsymbol{a}(\theta_{IRS})\boldsymbol{b}(\theta_{BS})$, where $\theta_{IRS}$ and $\theta_{BS}$ denote the angle of arrive (AoA) and the AoD of the BS-IRS link at the IRS and the BS, respectively. $\boldsymbol{a}(\theta_{IRS})$ denotes the steering vector at the IRS as (6), and $\boldsymbol{b}(\theta_{BS})$ denotes the steering vector at the BS as $\boldsymbol{b}(\theta_{BS}) = \left[1,e^{j\phi_l},\cdot\cdot\cdot,e^{j(N-1)\phi_l}\right]^T$ with $\phi_l = \frac{2\pi d_{BS}}{\lambda}sin\theta_{BS}$. Let $d_{BS}$ denote the spacing between consecutive antennas at the BS. Thus, the beampattern gain is rewritten as
%\begin{equation}
 %   \rho(\theta_l)= |\boldsymbol{\alpha}^H(\theta_{IRS})\boldsymbol{\Phi}^H\boldsymbol{\alpha}(\theta_l)|^2\boldsymbol{b}^H(\theta_{BS})\Big(\boldsymbol{R}_0 + \sum_{k=1}^K\boldsymbol{W}_k\Big)\boldsymbol{b}(\theta_{BS})
%\end{equation}

%In this condition, the dedicated radar signals are satisfied $\boldsymbol{R}_0^{opt} = 0$, meaning that the dedicated radar signals are not necessary for this condition, which has similar proofs in the \cite{r6}. 

%Second, we consider the problem (P2) when the cross-correlation pattern design between radar signals and the rank-one constraint are removed.
%\begin{proposition}\label{Prop.v3}
%In this condition, the optimal dedicated radar beamforming satisfies $R_0^{opt} = 0$. 
%\end{proposition}
%\begin{IEEEproof}
%See Appendix~\ref{proof-Prop.v3}.
%\end{IEEEproof}

\subsection{Optimization of IRS's Phase Shifting Matrix $\boldsymbol{\Phi}$}
In this subsection, we obtain the optimal phase shifting matrix $\boldsymbol{\Phi}$ of the IRS for problem (P2) under the given $(\{\boldsymbol{w}_k\}_{k\in{\cal K}},\boldsymbol{R}_0)$. We first define the following notations as
\begin{subequations}\label{eq.def2}
\begin{align}
 &\boldsymbol{F}_{k,j}^{\text{CU}} \triangleq \text{diag}(\boldsymbol{h}_{iu,k})^H\boldsymbol{G}\boldsymbol{R}_0\boldsymbol{G}^H\text{diag}(\boldsymbol{h}_{iu,j}),~\forall k,j\in{\cal K}\\
 &\boldsymbol{e}^{\text{CU}}_{k} \triangleq \text{diag}(\boldsymbol{h}_{iu,k})^H\boldsymbol{G}\boldsymbol{R}_0\boldsymbol{h}_{bu,k},~\forall k\in{\cal K},
\end{align}
\end{subequations}
Based on \eqref{eq.def2}, we define the augmented matrices as  $\bar{\boldsymbol{F}}_{k,j}^{\text{CU}}=[\boldsymbol{F}_{k,j}^{\text{CU}}, \boldsymbol{e}_{k}^{\text{CU}};(\boldsymbol{e}_{k}^{\text{CU}})^H,0]$, $\forall k, j\in{\cal K}$.

By introducing $\boldsymbol{V}\triangleq \bar{\boldsymbol{\phi}}\bar{\boldsymbol{\phi}}^H\succeq\boldsymbol{0}$ and $\text{rank}(\boldsymbol{V})\leq 1$, problem (P2) under the given $(\{\boldsymbol{w}_k\}_{k\in{\cal K}},\boldsymbol{R}_0)$ is recast as
\begin{subequations}
 \begin{align}
    (\text{P2.2}):&~\max_{\boldsymbol{V}}~  \min_{l\in \mathcal{L}}~ \text{tr}(\bar{\boldsymbol{R}}^{\text{target}}(\theta_l)\boldsymbol{V})\\
        \text{s.t.}~ &\text{tr}\Big(\Big(\frac{1}{\Gamma_k}\bar{\boldsymbol{R}}^{\text{CU}}_{k,k}-\sum_{j=1,j\neq k}^K \bar{\boldsymbol{R}}^{\text{CU}}_{k,j}-\bar{\boldsymbol{F}}^{\text{CU}}_{k}\Big)\boldsymbol{V}\Big) \geq c_k^{\text{II}},~ \forall k \in \mathcal{K}\\
        & (16c)-(16e), \nonumber
\end{align}
\end{subequations}
where $c_k^{\text{II}}\triangleq \sum_{j=1,j\neq k}^K |\boldsymbol{h}^H_{bu,k}\boldsymbol{w}_j|^2 -\frac{1}{\Gamma_k}|\boldsymbol{h}_{bu,k}^H\boldsymbol{w}_k|^2 + \sigma_k^2$, $\forall k \in{\cal K}$, and the matrix $\boldsymbol{V}$ is the design variable. As in problem (P1.2), problem (P2.2) can be relaxed into a SDP by employing the SDR technique. In the case with a high-rank solution for the relaxed problem (P2.2), a Gaussian randomization procedure is adopted to generate a feasible rank-one solution for problem (P2.2).

\subsection{Proposed Algorithm for (P2) with Type-II CUs}
In Algorithm~2, we summarize the proposed method for solving problem (P2) in the case with Type-II CUs. Based on the alternating-optimization-based method, we obtain a low-complexity solution to problem (P2) by alternatively solving problems (P2.1) and (P2.2) in an iteration manner. Since the objective function values are non-decreasing for problems (P2.1) and (P2.2), the convergence of Algorithm~2 is guaranteed. Similarly as Algorithm~1, the computational complexity of Algorithm~2 is given as $\mathcal{O}(J(K^{4.5}N^{4.5})\log(1/\epsilon)) + \mathcal{O}(JI(NK+1)^{2})$.

\begin{algorithm}
\caption{Proposed alternating-optimization algorithm for (P2)}
    \begin{algorithmic}[1]
    \STATE Initialize the phase shifting matrix $\boldsymbol{\Phi}^{(1)}$ and the iteration index $t = 1$;
    \REPEAT
    \STATE {For given $\Phi^{(t)}$, obtain the optimal solution $\boldsymbol{W}_k^{(t)}$ and $\boldsymbol{R}_0^{(t)}$ to problem (P2.1);}
    \STATE For given $\boldsymbol{W}_k^{(t+1)}$ and $\boldsymbol{R}_0^{(t+1)}$, solve the relaxed problem (P2.2), and then obtain a rank-one solution $\boldsymbol{\Phi}^{(t+1)}$ based on the Gaussian randomization method;
    \STATE  Set $t = t+1$;
     \UNTIL The objective value between two iterations is smaller than a tolerance level $\epsilon$ or the maximum number of iterations is reached.
    \end{algorithmic}
\end{algorithm}

\section{Numerical Results}
\begin{figure}
    \centering
    \includegraphics[width=12cm]{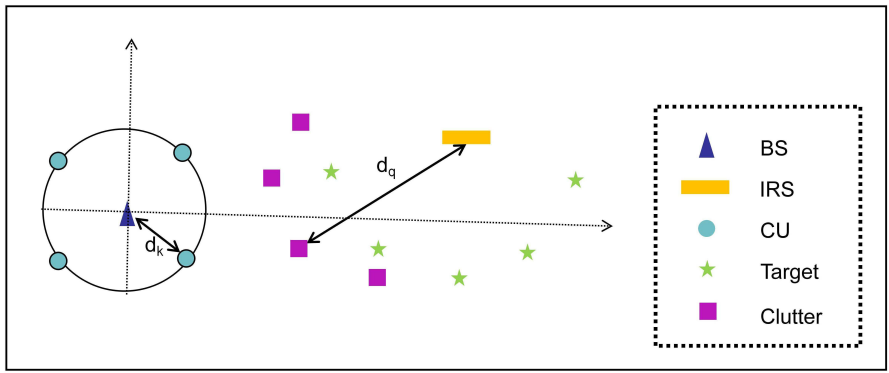}
    \caption{Simulation setup}\label{fig.d}
\end{figure}

 In this section, numerical results are provided to evaluate the proposed IRS-assisted ISAC system design scheme in a clutter environment. We consider the Rician fading channel models for the BS-CU, BS-IRS, BS-Clutter, IRS-CU, and IRS-Clutter communication links, where the Rician factor is set as 0.5 in the radar field. The pass-loss model is given by $K_0(\frac{d}{d_0})^{-\alpha}$, where $K_0 = -30 $ dB corresponds to the path-loss at the reference distance of $d_0 = 1$ meter (m), and the path-loss exponent $\alpha$ are set to be 2.5, 2.5, 2.2, 3.5, and 3.5 for the IRS-Clutter, IRS-CU, BS-IRS, BS-CU, and BS-Clutter links, respectively. The number of the BS's antennas is set to be $M=8$, and the number of the IRS's passive reflecting elements is set to be $N=64$. As shown in Fig.~\ref{fig.d}, the BS and IRS are set to be located at the coordinates (0 m, 0 m) and (20 m, 2 m), respectively. The number of CUs is $K = 4$, and each CU $k\in{\cal K}$ is randomly distributed with a distance $d_k=10$~m apart from the BS. Each clutter $q\in{\cal Q}$ is randomly located at a distance $d_q \in[12,15]$~m apart from the IRS. The angles of the $L=5$ targets are given as $\theta_1=-60^{\circ}$, $\theta_2=-30^{\circ}$, $\theta_3=0^{\circ}$, $\theta_4=30^{\circ}$, and $\theta_5=60^{\circ}$. Unless stated otherwise, the BS's transmission power budget is set to be $P_0 = 0.5$ Watt (W), the CU receiver noise power is set to be $\sigma_k^2 = -80$ dBm, and the interference power threshold of each clutter is set to be $\eta_q=0.1~\mu\text{W}$, $\forall q\in{\cal Q}$. Finally, the error tolerable level is set to be $\epsilon = 10^{-6}$ in the proposed Algorithms~1 and~2.

 For performance comparison, we consider the following four benchmark schemes for IRS-assisted ISAC system designs.
 \begin{itemize}
     \item {\em Information Beamforming Design Scheme:} In this scheme, the BS only employs the information beamforming vectors $\{\boldsymbol{w}_k\}_{k\in{\cal K}}$ for both communication and sensing, which is equivalent to solving problem \eqref{eq.p31} by setting $\boldsymbol{R}_0 = 0$.
     \item{\em Separate Design Scheme:} In this scheme, the BS's transmit design $(\{\boldsymbol{w}_k\}_{k\in{\cal K}},\boldsymbol{R}_0)$ and the IRS's phase shifting matrix $\boldsymbol{\Phi}$ are optimized separately. In particular, the phase shifting matrix $\boldsymbol{\Phi}$ at the IRS is first optimized so as to make the IRS reach the desired angle, and the transmission design $(\{\boldsymbol{w}_k\}_{k\in{\cal K}},\boldsymbol{R}_0)$ at the BS is then optimized under the the obtained phase shifting matrix $\boldsymbol{\Phi}$.
    \item {\em Joint Design with Random IRS Phase Scheme:} In this scheme, the IRS's phase shifting matrix $\boldsymbol{\Phi}$ is randomized generated, and then $(\{\boldsymbol{w}_k\}_{k\in{\cal K}},\boldsymbol{R}_0)$ is jointly optimized at the BS.
   \item {\em Joint Design without IRS Scheme:} In this scheme, the IRS design is removed, i.e., $\boldsymbol{\Phi} = 0$, and $(\{\boldsymbol{w}_k\}_{k\in{\cal K}},\boldsymbol{R}_0)$ at the BS becomes the only design variables for ISAC.
 \end{itemize}

\subsection{Convergence Performance}

\begin{figure}
\centering
\includegraphics[width=14cm]{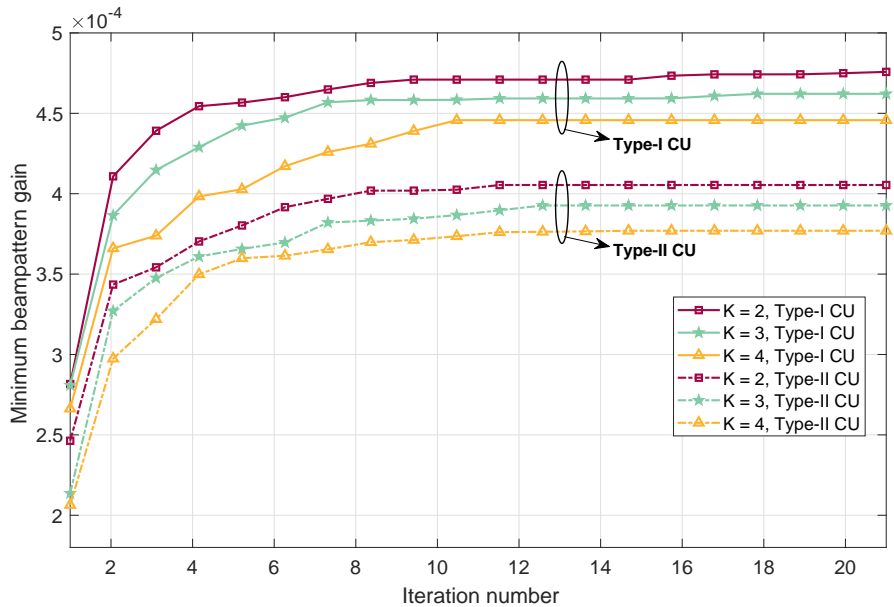}
\caption{Convergence performance of the proposed Algorithms~1 and 2.}\label{fig.conv}
\end{figure}

 Fig.~\ref{fig.conv} shows the fast convergence performance of the proposed Algorithms~1 and 2 for solving problems (P1) and (P2), respectively, where $Q = 2$, $L = 5$, $\xi = \infty$, and the SINR threshold for each CU $k\in{\cal K}$ is set to be $\Gamma_k = 10$~dB. It is observed that Algorithms~1 and 2 both converge to stationary points within about 17 iterations under different numbers of CUs. The convergence speed becomes slightly slow as the number $K$ of CUs increases. As expected, it is observed in Fig.~\ref{fig.conv} that the minimum sensing beampattern gain in the case with Type-I CUs is larger than that in the case with Type-II CUs. In addition, the achieved sensing beampattern gain becomes smaller as $K$ increases. This is because the max-min based ISAC system design scheme relies heavily on the CU with the worst channel quality so as to satisfy the communication and sensing requirements. 

\subsection{Achieved Sensing Beampattern Gain Profiles}

\begin{figure}
\centering
\includegraphics[width=13cm]{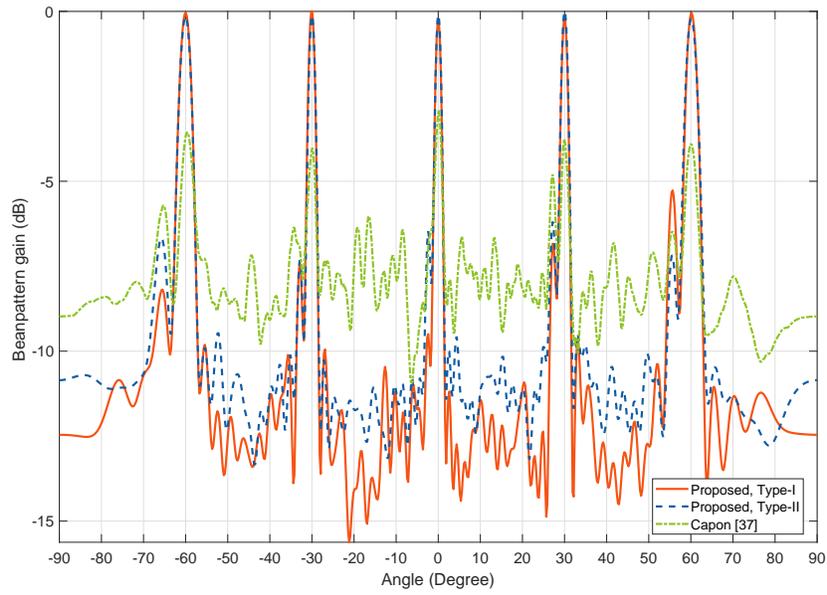}
\caption{The sensing beampattern gain profile for different schemes}\label{fig.diff}
\end{figure}

Fig.~\ref{fig.diff} depicts the sensing beampattern gain achieved at different angles, where $L = 5$, $K = 3$, $Q = 2$, and $\xi = \infty$. All three schemes concentrate energy at the sensing angles of the five targets. The proposed joint design schemes with different CU Types are observed to significantly outperform the conventional Capon based scheme [37]. Also, the proposed design in the case with Type-I CUs outperforms the proposed design in the case with Type-II CUs, i.e., a lower power profile at the uninterested angles is achieved in the case with Type-I CUs. This shows the benefits of cancelling the sensing interference at the CUs for improving the sensing performance. 

\begin{figure}
\centering
\includegraphics[width=13cm]{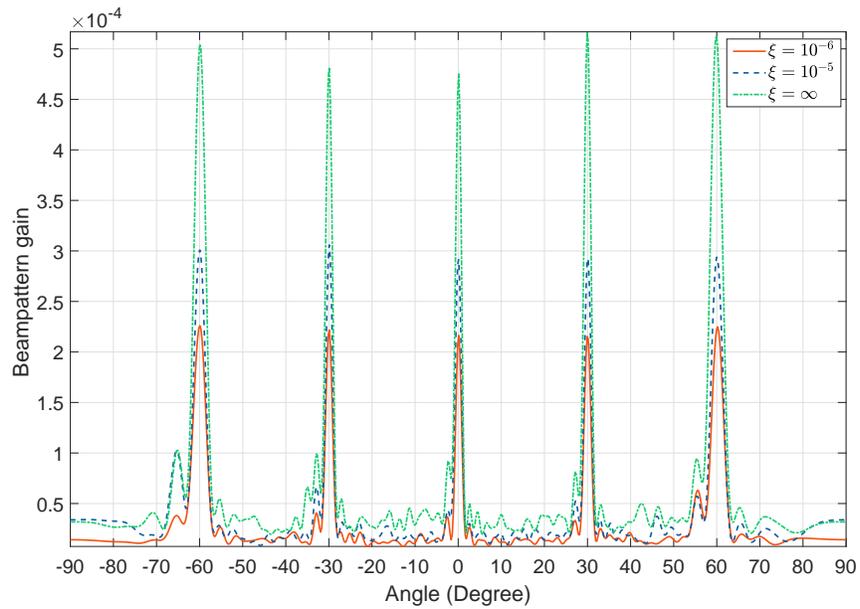}
\caption{The achieved sensing beampattern gain profile under different mean-squared cross-correlation threshold $\xi$ values.}\label{fig:cro}
\end{figure}

Fig.~\ref{fig:cro} shows the beampattern gain profile of the proposed scheme in the case with Type-I CUs under different sensing mean-squared cross-correlation pattern threshold $\xi$ values, where $K = 3$, $\Gamma_k= 10$~dB, $Q = 2$, and $L =5$. With a smaller threshold $\xi$, a stricter cross-correlation coefficient is imposed between the illuminating signals towards any two neighboring targets, which enhances the multi-target detection performance \cite{r44}. As the threshold $\xi$ value increases, the proposed design scheme is observed to achieve increasingly large sensing beampattern gains at the target angles. This implies the importance of properly setting a cross-correlation pattern threshold $\xi$, so as to achieve a good trade-off between the multi-target detection performance and the multi-target sensing performance \cite{r44}.

%\begin{figure}
%\centering
%\includegraphics[width=12cm]{figW.eps}
%\caption{The correlation coefficient between radar signal versus the sensing cross-correlation threshold $\xi$ with $L=4$.}\label{fig:xi}
%\end{figure}

%Fig.~\ref{fig:xi} shows the cross-correlation coefficient versus the sensing cross-correlation pattern threshold $\xi$ for the $L=5$ targets. The correlation coefficient between the illuminating signals towards any two neighboring targets is observed to linearly increases as the cross-pattern threshold $\xi$ decreases. When the sensing signals for illuminating different targets are highly correlated, it leads to an degradation for the multi-target detection and sensing performance \cite{r44}. 

\subsection{Achieved Minimum Sensing Beampattern Gain Evaluation}

\begin{figure}
    \centering
    \includegraphics[width=13cm]{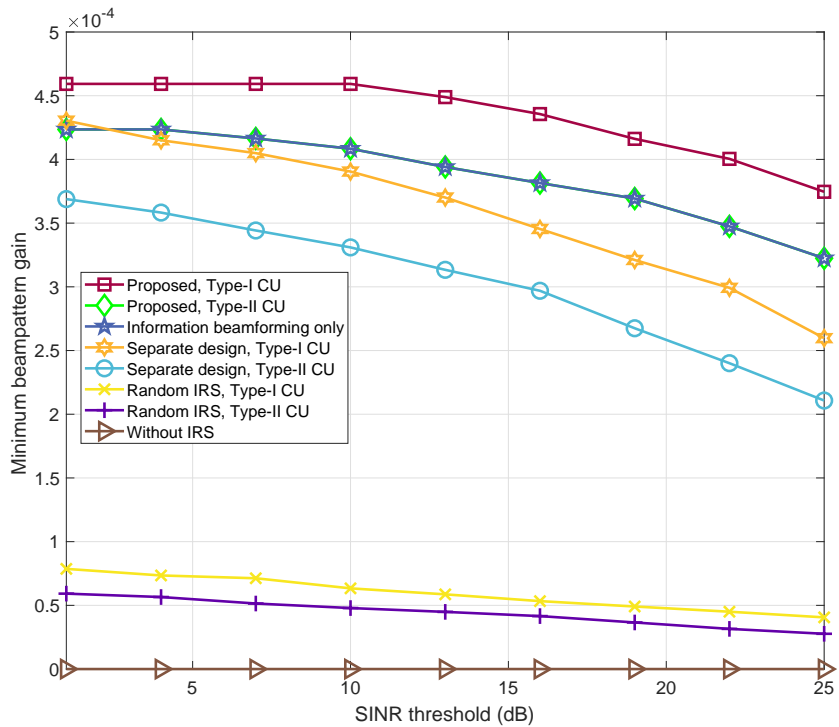}
    \caption{The achieved minimum beampattern gain versus the SINR threshold $\Gamma$ of CUs with $K = 3$, $Q = 2$, and $L = 5$.}\label{fig.SINR}
\end{figure}

Fig.~\ref{fig.SINR} shows the achieved minimum beampattern gain versus the SINR $\Gamma$ of the CUs, where $K = 3$, $Q = 2$, $L =5$, and $\xi = \infty$. The proposed designs are observed to achieve a significant performance gain over the four benchmark schemes. Thanks to the ability of eliminating the sensing signals for Type-I CUs, the proposed design scheme in the case with Type-I CUs outperforms the other schemes. In Fig.~\ref{fig.SINR}, the proposed scheme in the case with Type-II CUs is observed to achieve the same performance as the information beamforming design benchmark scheme, which corroborates Proposition~1. For the proposed design scheme in the case with Type-I CUs, the minimum beampattern gain remains unchanged in the low SINR requirement regime (e.g. $\Gamma \leq 10$~dB), but it is not true in the high SINR requirement regime (e.g. $\Gamma \geq 15$~dB). This shows the IRS-assisted ISAC system design needs to balance the trade-off between high communication and sensing performance requirements. In addition, the separate design benchmark scheme is observed to outperform the benchmark scheme with random IRS phase matrix. This implies the importance of optimizing the IRS phase matrix in improving the sensing performance. Finally, due to the fact that there exist no direct LoS links between the targets and the BS, the benchmark scheme without IRS is observed to fail to sense the targets. 
\begin{figure}
    \centering
    \includegraphics[width=12cm]{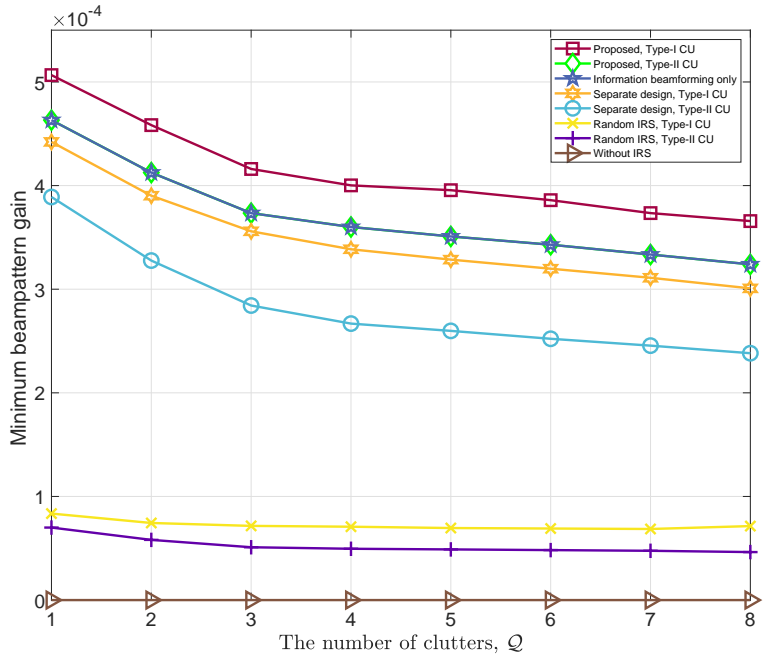}
    \caption{The achieved minimum beampattern gain versus the clutter number $Q$ with $K = 3$ and $\Gamma_k= 10$~dB.} \label{fig.clutter}
\end{figure}

Fig.~\ref{fig.clutter} shows the achieved minimum sensing beampattern gain versus the clutter number $Q$, where $K = 3$, $\Gamma_k= 10$~dB, $L =5$, and $\xi = \infty$. It is observed that the proposed schemes achieve a significant performance gain over the benchmark schemes. Apart from the benchmark schemes with random IRS or without the IRS, the minimum sensing beampattern gain decreases for the other schemes as the clutter number $Q$ increases. This is because a more amount of energy is needed to counter against the increasing clutter interference value. All the IRS-assisted benchmark schemes are observed to outperform the scheme without IRS. This is because the IRS can help create a sensing link to effectively illuminate the $L$ targets.

 \begin{figure}
    \centering
    \includegraphics[width=12cm]{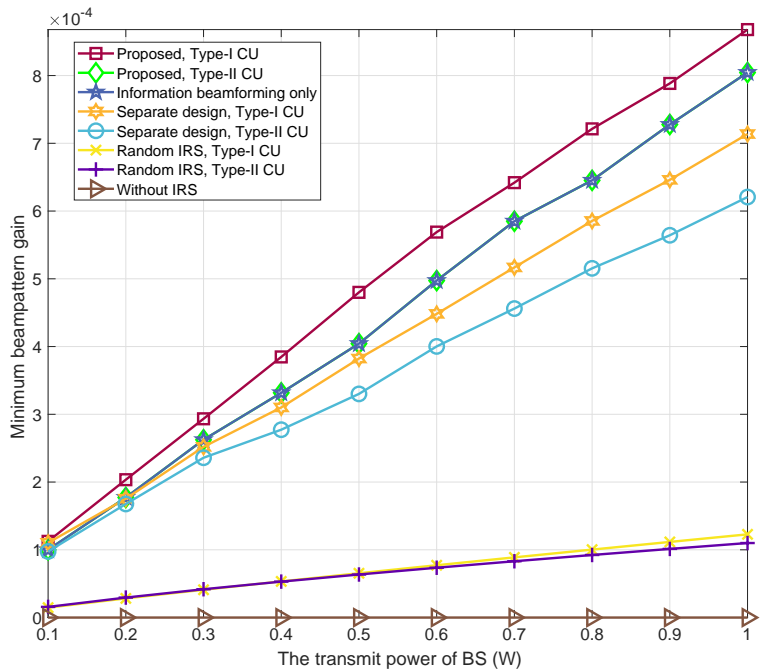}
    \caption{The achieved minimum sensing beampattern gain versus the BS's transmit power with $K = 3$, $Q = 2$, $\Gamma= 10$~dB and $N= 64$.}\label{fig:BS}
 \end{figure}

Fig.~\ref{fig:BS} shows the minimum sensing beampattern gain versus the maximum transmit power $P_0$ of BS, with $K = 3$, $\Gamma_k = 10$ dB, $Q = 2$, $L =5$, and $\xi = \infty$, respectively. Besides the benchmark scheme without IRS, the beampattern gain performance of the other schemes increases as the BS's maximum transmit power increases. Again, the proposed schemes are observed to outperform the other benchmark schemes. The proposed joint scheme in the case with Type-II CUs is also observed to have the same performance as the information beamforming design benchmark scheme. The design scheme with a random IRS phase matrix performs inferiorly to the other schemes with an optimized IRS phase matrix.

\begin{figure}
    \centering
    \includegraphics[width=12cm]{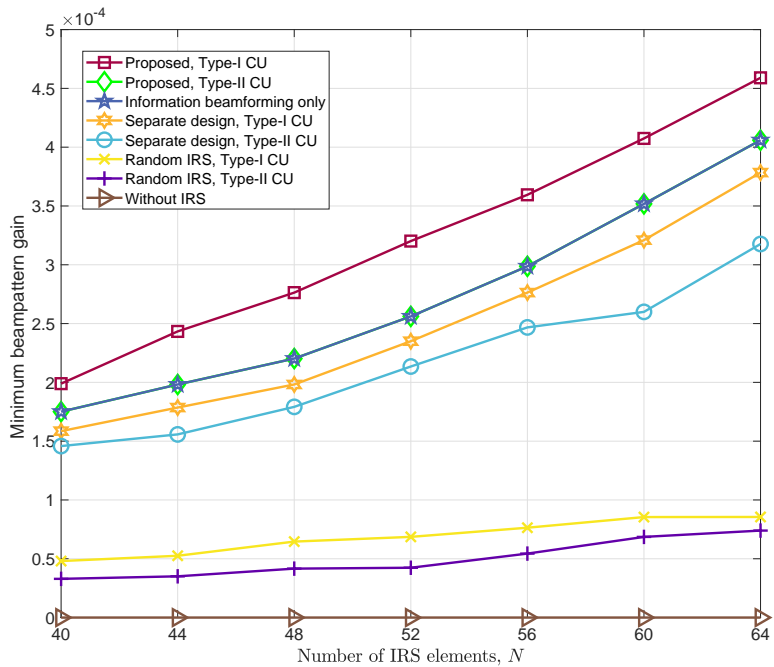}
    \caption{The minimum sensing beampattern gain versus the number $N$ of the IRS passive reflecting elements with $K = 3$, $Q = 2$, and $\Gamma = 10$~dB.}\label{fig:N}
\end{figure}

Fig.~\ref{fig:N} shows the achieved minimum sensing beampattern gain versus the number $N$ of the IRS passive reflecting elements, where $K = 3$, $\Gamma_k= 10$~dB, $Q = 2$, $L =5$, and $\xi = \infty$. As expected, the achieved minimum sensing beampattern gain of all the schemes increases as $N$ increases. This is because the IRS with a larger number of passive reflecting elements is able to provide a higher degrees of freedom to reconfigure the signal propagation environment, thereby improving the system sensing performance. The proposed scheme in the case with Type-I CUs is observed to achieve a significant performance gain than the other schemes, and the proposed scheme in the case with Type-II CUs achieves the same performance as the information beamforming only design scheme.

\section{Conclusion}
In this paper, we studied an IRS-assisted ISAC system design in a clutter environment, where the IRS is deployed to both assist the performance of multiuser communication and multi-target sensing. We consider the Type-I CUs and Type-II CUs based on the ability of eliminating the sensing interference for CUs. In the case with Type-I or Type-II CUs, we maximized the minimum beampattern gain among multiple targets by jointly optimizing the BS's transmit beamforming vectors for multiple CUs and covariance matrix for sensing multiple targets, as well as the IRS's passive phase shifting matrix. For the formulated non-convex max-min ISAC design problems in the cases with Type-I CUs and Type-II CUs, we developed low-complexity algorithms based on the alternating-optimization and SDR techniques. It is shown that the dedicated sensing signals for targets are necessary to improve the system performance in the case with Type-I CUs. By contrast, the dedicated sensing signals can be removed for the ISAC system designs in the case with Type-II CUs. Numerical results demonstrated that the proposed IRS-assisted ISAC design in the case with Type-I CUs outperforms that in the case with Type-II Cus, and both the proposed joint design schemes achieve a significant performance gain than the existing benchmark schemes.

\appendix
\subsection{Proof of Lemma~\ref{Prop.v1}}\label{proof-Prop.v1}

First, it follows from (\ref{eq.lem1}b) that $\tilde{\boldsymbol{R}}^{\text{I}*}_0+ \sum_{k=1}^{K}\tilde{\boldsymbol{w}}^{\text{I}*}_k(\tilde{\boldsymbol{w}}^{\text{I}*}_k)^H =\boldsymbol{R}_0^{\text{I}*}+\sum_{k=1}^{K}\boldsymbol{W}_k^{\text{I}*}$. Therefore, the solution of $(\{\tilde{\boldsymbol{W}}^{\text{I}*}_k\}_{k\in{\cal K}}, \tilde{\boldsymbol{R}}^{\text{I}*}_0)$ satisfies the constraints (\ref{eq.p1-1}b), (\ref{eq.p1-1}d), and (\ref{eq.p1-1}e). In addition, the objective function value of problem (P1.1) achieved by $(\{\tilde{\boldsymbol{W}}^{\text{I}*}_k\}_{k\in{\cal K}}, \tilde{\boldsymbol{R}}^{\text{I}*}_0)$ is exactly the same as that achieved by $(\{\boldsymbol{W}^{\text{I}*}_k\}_{k\in{\cal K}}, \boldsymbol{R}^{\text{I}*}_0)$. 

Next, we show the solution of $\{\tilde{\boldsymbol{W}}^{\text{I}*}_k\}$ satisfies the constraint (\ref{eq.p1-1}c). Based on (\ref{eq.lem1}a), we have $\tilde{\boldsymbol{w}}^{\text{I}*}_k= \frac{1}{\sqrt{\boldsymbol{h}^H_{\text{CU},k}\boldsymbol{W}_k^{\text{I}*}\boldsymbol{h}_{\text{CU},k}}} \boldsymbol{W}_k^{\text{I}*}\boldsymbol{h}_{\text{CU},k}$, and $\tilde{\boldsymbol{W}}^{\text{I}*}_k=\tilde{\boldsymbol{w}}^{\text{I}*}_k(\tilde{\boldsymbol{w}}^{\text{I}*}_k)^H$, $\forall k\in{\cal K}$. For an arbitrary vector $\boldsymbol{y}\in \mathbb{C}^M$, it follows that
\begin{equation}\label{eq.a1}
    \boldsymbol{y}^H\left(\boldsymbol{W}^{\text{I}*}_k-\tilde{\boldsymbol{W}}^{\text{I}*}_k\right)\boldsymbol{y}=\boldsymbol{y}^H\boldsymbol{W}^{\text{I}*}_k\boldsymbol{y}-\left(\boldsymbol{h}^H_{\text{CU},k}\boldsymbol{W}^{\text{I}*}_k\boldsymbol{h}_{\text{CU},k}\right)^{-1}|\boldsymbol{y}_k^H\boldsymbol{W}^{\text{I}*}_k\boldsymbol{h}_{\text{CU},k}|^2, \forall k\in \mathcal{K}.
\end{equation}
In addition, from the Cauchy-Schwarz inequality, it yields that 
\begin{equation}\label{eq.a2}
    \left(\boldsymbol{y}^H\boldsymbol{W}^{\text{I}*}_k\boldsymbol{y}\right)\left(\boldsymbol{h}^H_{\text{CU},k}\boldsymbol{W}^{\text{I}*}_k\boldsymbol{h}_{\text{CU},k}\right)\geq\left|\boldsymbol{y}_k^H\boldsymbol{W}^{\text{I}*}_k\boldsymbol{h}_{\text{CU},k}\right|^2, \forall k\in \mathcal{K}.
\end{equation}
Based on \eqref{eq.a1} and \eqref{eq.a2}, we have
\begin{equation}
     \boldsymbol{y}^H\left(\boldsymbol{W}^{\text{I}*}_k-\tilde{\boldsymbol{W}}^{\text{I}*}_k\right)\boldsymbol{y}\geq0\Longleftrightarrow\boldsymbol{W}^{\text{I}*}_k-\tilde{\boldsymbol{W}}^{\text{I}*}_k\succeq0, \forall k\in \mathcal{K}.
\end{equation}

Furthermore, it holds that $\boldsymbol{h}^H_{\text{CU},k}\tilde{\boldsymbol{W}}^{\text{I}*}_k\boldsymbol{h}_{\text{CU},k}=\boldsymbol{h}^H_{\text{CU},k}\tilde{\boldsymbol{w}}^{\text{I}*}_k\tilde{\boldsymbol{w}}^{\text{I}*}_k\boldsymbol{h}_{\text{CU},k}=\boldsymbol{h}^H_{\text{CU},k}\boldsymbol{W}^{\text{I}*}_k\boldsymbol{h}_{\text{CU},k}$, $\forall k\in \mathcal{K}$. By expressing (\ref{eq.p1-1}c) as $(1+\frac{1}{\Gamma_j})\boldsymbol{h}^H_{\text{CU},j}\boldsymbol{W}_j\boldsymbol{h}_{\text{CU},j} -\boldsymbol{h}^H_{\text{CU},j}(\sum_{k=1}^{K}\boldsymbol{W}_k)\boldsymbol{h}_{\text{CU},j}-\sigma_j^2\geq 0$, $\forall k, j\in \mathcal{K}$, we are ready to have
\begin{subequations}
    \begin{align}   \Big(1+\frac{1}{\Gamma_j}\Big)\boldsymbol{h}^H_{\text{CU},j}\tilde{\boldsymbol{W}}^{\text{I}*}_j\boldsymbol{h}_{\text{CU},j}&=
        \Big(1+\frac{1}{\Gamma_j}\Big)\boldsymbol{h}^H_{\text{CU},j}\boldsymbol{W}^{\text{I}*}_j
        \boldsymbol{h}_{\text{CU},j}\\
        &\geq\boldsymbol{h}^H_{\text{CU},j}\Big(\sum_{k=1}^{K}\boldsymbol{W}^{\text{I}*}_k
        \Big)\boldsymbol{h}_{\text{CU},j}+\sigma_j^2 \\
        &\geq\boldsymbol{h}^H_{\text{CU},j}\Big(\sum_{k=1}^{K}\tilde{\boldsymbol{W}}^{\text{I}*}_k\Big)\boldsymbol{h}_{\text{CU},j}+\sigma_j^2,~\forall k, j\in{\cal K}.
    \end{align}
\end{subequations}

Therefore, the reconstructed rank-one solution  $(\{\tilde{\boldsymbol{W}}^{\text{I}*}_k\}_{k\in{\cal K}}, \tilde{\boldsymbol{R}}^{\text{I}*}_0)$ satisfies the SINR constraints of Type-I CUs in (\ref{eq.p1-1}c). Until now, we have proved that the solution $(\{\tilde{\boldsymbol{W}}^{\text{I}*}_k\}_{k\in{\cal K}}, \tilde{\boldsymbol{R}}^{\text{I}*}_0)$ is an optimal solution to problem (P1.1), which completes the proof of Lemma~\ref{Prop.v1}.

\subsection{Proof of Proposition~\ref{Prop.v2}}\label{proof-Prop.v2}
First, by setting $\boldsymbol{R}_0=\boldsymbol{0}$ in the relaxed problem (P2.1), we obtain the following optimization problem as
\begin{subequations}\label{eq.p31}
 \begin{align}
 \max_{\{\boldsymbol{W}_k\succeq \boldsymbol{0}\}}~  \min_{l\in \mathcal{L}}~&~ \boldsymbol{h}_{\text{target},l}^H\Big(\sum_{k=1}^K\boldsymbol{W}_k\Big)\boldsymbol{h}_{\text{target},l}\\
  \text{s.t.}~ &~\sum_{k=1}^K\text{tr}(\boldsymbol{W}_k)\leq P_0 \\
  &~\frac{1}{\Gamma_k}\text{tr}(\boldsymbol{h}_{\text{CU},k} \boldsymbol{h}_{\text{CU},k}^H \boldsymbol{W}_k) - \sum_{j=1,j\neq k}^K\text{tr}(\boldsymbol{h}_{\text{CU},k}\boldsymbol{h}_{\text{CU},k}^H\boldsymbol{W}_j)-\sigma_k^2\geq 0,~\forall k \in \mathcal{K}\\
  &~\boldsymbol{g}_{\text{clutter},q}^H\Big(\sum_{k=1}^K\boldsymbol{W}_k \Big) \boldsymbol{g}_{\text{clutter},q} \leq \eta_q,~ \forall q \in \cal{Q}\\
  &~\frac{2}{L^2-L}\sum_{l=1}^{L-1}\sum_{i=l+1}^{L} \boldsymbol{h}_{\text{target},l}^H\Big(\sum_{k=1}^K\boldsymbol{W}_k\Big)\boldsymbol{h}_{\text{target},i}^H \leq \xi
\end{align}
\end{subequations}
Denote by $(\{\boldsymbol{W}^*_k\}_{k\in{\cal K}})$ the optimal solution to problem \eqref{eq.p31}. In particular, it holds that $\boldsymbol{W}^*_k = \boldsymbol{W}^{\text{II*}}_k + \beta_k\boldsymbol{R}^{\text{II*}}_0$, $\forall k\in{\cal K}$, where $\sum_{k=1}^K\beta_k = 1$ and $\beta_k \geq 0$, $\forall k\in{\cal K}$.

Next, we prove that $(\{\boldsymbol{W}^*_k\}_{k\in{\cal K}},\boldsymbol{0})$ is an optimal solution to problem (P2.1). For each Type-II CU $j\in{\cal K}$, it follows that 
\begin{subequations}
    \begin{align}
    \Big(1+\frac{1}{\Gamma_j}\Big)\boldsymbol{h}^H_{\text{CU},j}(\boldsymbol{W}^*_j +\boldsymbol{0})\boldsymbol{h}_{\text{CU},j} &=  \Big(1+\frac{1}{\Gamma_j}\Big)\boldsymbol{h}^H_{\text{CU},j}(\boldsymbol{W}^{\text{II*}}_j + \beta_k\boldsymbol{R}^{\text{II}*}_0)\boldsymbol{h}_{\text{CU},j}\\
    &\geq \Big(1+\frac{1}{\Gamma_j}\Big)\boldsymbol{h}^H_{\text{CU},j}(\boldsymbol{W}^{\text{II*}}_j)\boldsymbol{h}_{\text{CU},j}\\
    &\geq \boldsymbol{h}^H_{\text{CU},j}\Big(\sum_{k=1}^{K}\boldsymbol{W}^{\text{II*}}_k + \boldsymbol{R}^{\text{II}*}_0\Big)\boldsymbol{h}_{\text{CU},j} + \sigma_j^2 \\
    &= \boldsymbol{h}^H_{\text{CU},j}\Big(\sum_{k=1}^{K}\boldsymbol{W}^*_k+ \boldsymbol{0}\Big)\boldsymbol{h}_{\text{CU},j} + \sigma_j^2,~\forall k, j\in{\cal K},
    \end{align}
\end{subequations}
where the first inequality holds from the fact $\boldsymbol{R}^{\text{II}*}_0\succeq 0$, and the second inequality holds from that the SINR constraint for Type-II CU $j$. Therefore, $(\{\boldsymbol{W}^*_k\}_{k\in{\cal K}},\boldsymbol{0})$ is shown to be a feasible solution to problem (P2.1).

Finally, it is readily to show that both $(\{\boldsymbol{W}^*_k\}_{k\in{\cal K}},\boldsymbol{0})$ and $(\boldsymbol{W}^{\text{II*}}_k,\boldsymbol{R}^{\text{II*}}_0)$ can achieve the same objective function value for problem (P2.1). Now, we have proved that $(\{\boldsymbol{W}^*_k\}_{k\in{\cal K}},\boldsymbol{0})$ is an optimal solution to problem (P2.1), and we complete the proof of Proposition~\ref{Prop.v2}. 

\bibliographystyle{IEEEtran}
\bibliography{ref}

\end{document}